\documentstyle[psfig]{siamltex}
\makeatletter
\let\oldnewcommand=\newcommand
\def\newcommand#1{\@ifnextchar [{\jhhnc#1}{\jhhnc#1[0]}}
\def\jhhnc#1[#2]{\@ifnextchar [{\jhhoptpar#1[#2]}{\oldnewcommand#1[#2]}}
\def\jhhoptpar#1[#2][#3]#4{%
  \expandafter\oldnewcommand\csname\string#1x\endcsname[#2]{#4}%
  \expandafter\def\csname\string#1\endcsname[##1]%
	{\csname\string#1x\endcsname{##1}}%
  \oldnewcommand#1{\@ifnextchar [{\csname\string#1\endcsname}%
                                 {\csname\string#1\endcsname[#3]}}
}
\makeatother

\renewcommand{\term}[1]{{\em #1\/}}	
\newcommand{\ie}{{i.e.},\ }		
\newcommand{\edge}[2]{({#1},{#2})}			
\newcommand{\degree}[2][{}]{d_{#1}({#2})} 	
\newcommand{\Set}[2]{\left\{{#1},\ldots,{#2}\right\}} 
\newcommand{\fac}[1]{{#1} !}		
\newcommand{\ceil}[1]{\lceil{#1}\rceil} 
\newtheorem{cla}{\sc Claim}
\newenvironment{claim}{\begin{cla}}{\end{cla}}

\title{Space-Efficient Routing Tables for Almost All Networks
 and the Incompressibility Method\thanks{A preliminary version of part
of this work was presented at the %
{\it 15th ACM Conf. Principles Distribut. Comput.,
Philadelphia, PA, USA, May 1996}.
All
authors were partially supported by the European Union
through NeuroCOLT ESPRIT Working Group Nr. 8556,
and by  NWO through NFI Project ALADDIN
number NF 62-376.
}}

\author{Harry Buhrman\thanks{CWI,
Kruislaan 413, 1098 SJ Amsterdam, The Netherlands; Email:
{\tt buhrman@cwi.nl}.}
\and
Jaap-Henk Hoepman\thanks{KPN Research,
P.O. Box 15000, 9700 CD Groningen, The Netherlands;
Email: {\tt J.H.Hoepman@research.kpn.com}}
\and
Paul Vit\'{a}nyi\thanks{CWI and University of Amsterdam. Address: CWI,
Kruislaan 413, 1098 SJ Amsterdam, The Netherlands;
Email: {\tt paulv@cwi.nl}}
}

\date{}

\newcommand{\mor}{$\vee$ }
\newcommand{\mand}{$\wedge$ }

\begin{document}
\maketitle

\pagestyle{myheadings}
\thispagestyle{plain}
\markboth{H.~M. BUHRMAN, J.~H. HOEPMAN, AND P.~M.~B. VIT\'ANYI}{COMPACT ROUTING TABLES}

\bibliographystyle{plain}

\begin{abstract}
We use the incompressibility method based on Kolmo\-go\-rov complexity
to determine the total number of bits
of routing information for almost all network topologies.
In most models for routing, for almost all labeled graphs
$\Theta (n^2)$ bits
are necessary and sufficient for shortest path routing.
By `almost all graphs' we mean the
Kolmogorov random graphs which constitute a fraction of $1-1/n^c$
of all graphs on $n$ nodes,
where $c > 0$ is an arbitrary fixed constant.
There is a model for which the average case lower bound rises to
$\Omega(n^2 \log n )$ and another model where the average case upper bound
drops  to $O(n \log^2 n)$.
This clearly exposes the sensitivity of such bounds to the model under
consideration.
If paths have to be short, but need not be shortest (if the
stretch factor may be larger than $1$), then
much less space is needed on average, even in the more demanding models.
Full-information routing requires $\Theta (n^3)$ bits on average.
For worst-case static networks we prove a
$\Omega(n^2 \log n )$ lower bound for shortest path routing
and all stretch factors $<2$ in some networks
where free relabeling is not allowed.
\end{abstract}

\section{Introduction}
In very large communication networks like the global
telephone network or the internet connecting the worlds
computers, the message volume being routed
creates bottlenecks degrading
performance.
We analyze a tiny part of this issue
by determining the optimal space to
represent routing schemes in
communication networks
for allmost all static network topologies. The results give
also the average space cost over all network topologies.

A universal \term{routing strategy} for static communication networks will,
for every network, generate a \term{routing scheme} for that particular
network. Such a routing scheme comprises a \term{local routing function} for
every node in this network.
The routing function of node $u$ returns for every destination $v \neq u$ an
edge incident to $u$ on a path from $u$ to $v$.
This way, a routing scheme describes a
path, called a \term{route}, between every pair of nodes $u,v$ in the network.
The \term{stretch factor} of a routing scheme equals the maximum ratio
between the length of a route it produces, and the shortest path between
the endpoints of that route.

It is easy to see that we can do shortest path
routing by entering a routing table in each
node $u$ which for each destination node $v$ indicates
to what adjacent node $w$ a message to $v$
should be routed first. If $u$ has degree $d$,
it requires a table of at most $n \log d$ bits\footnote{Throughout,
``$\log$'' denotes the binary logarithm.}
and the overall number of bits in all local
routing tables never exceeds $n^2 \log n$.

The stretch factor of a routing strategy equals
the maximal stretch factor attained by any of the routing schemes it generates.
If the stretch factor of
a routing strategy equals $1$, it is called a
\term{shortest path routing strategy}
because then it generates for every graph a routing scheme that will route a
message between arbitrary $u$ and $v$ over a shortest path between $u$ and $v$.

In a \term{full information} shortest path routing scheme,
the routing function in $u$ must, for each destination $v$
return all edges incident to $u$ on shortest paths from $u$ to $v$.
These schemes allow alternative, shortest, paths to be taken whenever an
outgoing link is down.

We consider point to point communication networks on $n$ nodes described
by an undirected graph $G$. The nodes of the graph initially have unique
labels taken from a set $\Set{1}{m}$ for some $m>n$. 
Edges incident to a node $v$ with degree
$\degree{v}$ are connected to \term{ports}, with fixed labels
$1,\ldots,\degree{v}$, by a so called \term{port assignment}.
This labeling corresponds to the minimal local
knowledge a node needs to route: a) a unique identity to determine whether it
is the destination of an incoming message, b) the guarantee that each of its
neighbours can be reached over a link connected to exactly one of its ports,
and c) that it can distinguish these ports.

\subsection{Cost Measures for Routing Tables}
The space requirements of a routing scheme is measured as the sum over all
nodes of the number of bits needed on each node to encode its
routing function. If the nodes are not labeled with
$\Set{1}{n}$---the minimal set of labels---we have to add
to the space
requirement, for each node, the number of bits needed to encode its label.
Otherwise, the bits needed to represent the routing function could be appended
to the original identity yielding a large label that is not charged for but
does contain all necessary information to route.

The cost of representing a routing function at a particular node depends on the
amount of (uncharged) information initially there.
Moreover, if we are allowed to relabel the graph and change its port assignment
before generating a routing scheme for it, the resulting routing functions may
be simpler and easier to encode.
On a chain, for example,
the routing function is much less complicated if
we can relabel the graph and number
the nodes in increasing order along the chain.
We list these assumptions below, and argue that each of them is reasonable for
certain systems. We start with the three options for
the amount of information initially available at a node.
\begin{itemize}
\item[I]
  Nodes do not initially know the labels of their neighbours, and use ports
  to distinguish the incident edges. This models the basic system without
  prior knowledge.
  \begin{itemize}
  \item[IA]
    The assignment of ports to edges is fixed and cannot be altered. This
    assumption is reasonable for systems running several jobs where
    the optimal port assignment for routing may actually be bad for those
    other jobs.
  \item[IB]
    The assignment of ports to edges is free and can be altered
    before computing the routing scheme
    (as long as neighbouring nodes remain neighbours after re-assignment).
    Port re-assignment is justifiable as a local action that usually can be
    performed without informing other nodes.
  \end{itemize}
\item[II]
  Nodes know the labels of their neighbours,
  and know over which edge to reach them. This information is for free.
  Or, to put it another way, an incident edge carries the same label as the
  node it connects to.
  This model is concerned only with the additional cost of routing
  messages beyond the immediate neighbours, and applies to systems where the
  neighbours are already known for various other reasons
  \footnote{We do not consider models that give neighbours for free
  and, at the same time, allow free port assignment. For, given a labeling
  of the edges by the nodes they connect to, the
  actual port assignment doesn't matter at all, and can in fact
  be used to represent $\degree{v} \log \degree{v}$ bits of the routing
  function. Namely, each assignment of ports corresponds to a permutation
  of the ranks of the neighbours --- the neighbours at port $i$ moves to
  position $i$. There are $\fac{\degree{v}}$ such permutations.}.
\end{itemize}
Orthogonal to that, the following three options regarding the labels of the
nodes are distinguished.
\begin{itemize}
\item[$\alpha$]
  Nodes cannot be relabeled. For large scale distributed systems relabeling
  requires global coordination that may be undesirable or simply impossible.
\item[$\beta$]
  Nodes may be relabeled before computing the routing scheme, but the range of
  the labels must remain $1,\ldots,n$.
  This model allows a bad distributions of labels to be avoided.
\item[$\gamma$]
  Nodes may be given arbitrary labels before computing the routing scheme,
  but the number of bits used to store its label is added to the space
  requirements of a node. Destinations
  are given using the new, complex, labels%
  \footnote{In this model it is assumed that a routing function cannot tell
  valid from invalid labels, and that a routing function always receives
  a valid destination label as input. Requiring otherwise makes the problem
  harder.}.
  This model allows us to store additional routing information, e.g.\
  topological information, in the label of a node.
  This sort of network may be appropriate
  for centrally designed interconnect networks for multiprocessors
  and communication networks. A common example architecture of this
type is the binary $n$-cube
network where the $2^n$ nodes are labeled with elements of $\{0,1\}^n$
such that there is an edge between each pair of nodes 
iff their labels differ in exactly one bit position. In this case one can 
shortest path route using only the labels by successively traversing
edges corresponding to flipping successive bits in the 
positions where source node and destination node differ.
\end{itemize}
These two orthogonal sets of assumptions IA, IB, or II, and
$\alpha$, $\beta$, or $\gamma$, define the
nine different models we will consider in this paper.
We remark that the lower bounds for models without relabeling are
less surprising and less hard to prove than the bounds for the
other models.

\subsection{Outline}

We determine the optimum space used to represent
shortest path routing schemes
on almost all labeled graphs, namely the 
Kolmogorov random graphs with randomness deficiency at most $c \log n$
which constitute
a fraction of at least $1-1/n^c$ of all graphs, for every fixed
constant $c>0$.
These bounds straightforwardly imply the same bounds for
the average case over all graphs provided we choose 
$c \geq 3$.
For an overview of the results, refer
to Table~\ref{tab.results}\footnote{In this table, arrows indicate that the
bound for that particular model follows from the bound found by tracing the
arrow. In particular, the average case lower bound for model
IA \mand $\beta$ is the same as the IA \mand $\gamma$ bound found
by tracing $\rightarrow$.
The reader may have guessed that a ? marks an open question}.

\begin{table*}
\footnotesize{
\begin{center}
\begin{tabular}{lccc}
        &{\it no relabeling\/}
		&{\it permutation\/}
			& {\it free relabeling\/} \\
	&  ($\alpha$) &  ($\beta$) & ($\gamma$) \\
\hline\hline
\multicolumn{4}{l}{{\bf worst case} --- {\it lower bounds}} \\
{\it port assignment free} (IB)
	& $\rightarrow$
		& $\Omega (n^2 \log n)$~\cite{GavP96}
			& $n^2/32$ [Thm~\ref{theo.lb1-a}] \\
{\it neighbours known} (II)
	& $(n^2/9) \log n$ [Thm~\ref{theo.aglb2}]  & $\Omega(n^2)$~\cite{FraG95b}
		 & $\Omega(n^{7/6})$~\cite{PelU89} \\
\hline\hline
\multicolumn{4}{l}{{\bf average case} --- {\it upper bounds}} \\
{\it port assignment fixed} (IA)
	& $(n^2/2) \log n$ [Thm~\ref{theo.ub6}]& $\leftarrow$ & $\leftarrow$ \\
{\it port assignment free} (IB)
	& $3n^2$ [Thm~\ref{theo.ub1}] & $\leftarrow$ & $\leftarrow$ \\
{\it neighbours known} (II)
	& $3n^2$ [Thm~\ref{theo.ub1}]& $\leftarrow$  & $6n \log^2 n$ [Thm~\ref{theo.ub5}]  \\
\hline\hline
\multicolumn{4}{l}{{\bf average case} --- {\it lower bounds}} \\
{\it port assignment fixed} (IA)
	& $(n^2/2) \log n$ [Thm~\ref{theo.lb2}] & $\rightarrow$ &  $n^2/32$ [Thm~\ref{theo.lb1-a}] \\
{\it port assignment free} (IB)
	&  $n^2/2$ [Thm~\ref{theo.lb1}] & $\rightarrow$ & $n^2/32$ [Thm~\ref{theo.lb1-a}] \\
{\it neighbours known} (II)
	& $n^2/2$ [Thm~\ref{theo.lb1}] & ? & ? \\
\end{tabular}
\end{center}
\caption{Size of shortest path routing schemes: overview of results. The results
presented in this paper are quoted with exact constants and asymptotically
(with the lower order of magnitude terms suppressed).}
\label{tab.results}
}
\end{table*}

We prove that for almost all graphs
$\Omega(n^2)$ bits are necessary to represent
the routing scheme, if relabeling is not allowed and
nodes know their neighbours (II \mand $\alpha$) or nodes
do not know their neighbours (IA \mor IB)\footnote{We write A \mor B to
indicate that the results hold under model A or model B. Similarly, we write
A \mand B to indicate the result holds only if the conditions of both model A
and model B hold simultaneously. If only one of the two `dimensions' is
mentioned, the other may be taken arbitrary (\ie IA is a shorthand for
(IA \mand $\alpha$) \mor (IA \mand $\beta$) \mor (IA \mand $\gamma$)).}.
Partially matching this lower bound, we show that
$O(n^2)$ bits are sufficient to represent
the routing scheme, if the port assignment may be changed or
if nodes do know their neighbours (IB \mor II).
In contrast, for almost all graphs, the lower bound rises
to asymptotically $n^2 /2 \log n$ bits
if both relabeling and changing the port
assignment are not allowed (IA \mand $\alpha$), and this 
number of bits is also sufficient for almost all graphs.
And, again for almost all graphs,
the upper bound drops to $O(n \log^2 n)$ bits
if nodes know the labels of their neighbours and nodes may be
arbitrarily relabeled (II \mand $\gamma$).

Full information shortest path routing schemes are shown to require,
on almost all graphs, asymptotically $ n^3 /4 $ bits to be stored, if
relabeling is not allowed ($\alpha$), and this number of bits
is also shown to be sufficient for almost all graphs.
(The obvious upper bound for all graphs is $n^3$ bits.)

For stretch factors larger than $1$ we obtain the following results.
When nodes know their neighbours (II),
for almost all graphs, routing schemes achieving stretch factors $s$ with
$1 < s < 2$ can be stored using a total of $O(n \log n)$ bits%
\footnote{For
Kolmogorov random graphs which have diameter 2 by Lemma~\ref{lem.diam}
routing schemes with $s = 1.5$ are the only ones possible in this range.}.
Similarly, for almost all graphs in the same models (II),
$O (n \log \log n )$ bits are sufficient
for routing with stretch factor $\geq 2$.
Finally, for stretch factors $\geq 6 \log n$ on almost
all graphs again in the same model (II), the routing scheme occupies only
$O(n)$ bits.

For worst case static networks we prove, by construction of explicit graphs,
a $\Omega (n^2\log n)$ lower bound on the total size of
any routing scheme with stretch factor $< 2$, if
nodes may not be relabeled ($\alpha$).

The novel incompressibility technique based on Kolmogorov
complexity, \cite{LiV93}, has already been applied in many areas
but not so much in a distributed setting.
A methodological contribution of this paper is to show how to apply the 
incompressibility method to obtain
results in distributed computing for {\em almost all} objects concerned,
rather than for the {\em worst-case} object. This hinges
on our use of Kolmogorov random graphs in a fixed family of graphs. Our results
hold also {\em averaged over all} objects concerned.

Independent recent work
\cite{KraKU95b,KraK95} applies Kolmogorov
complexity to obtain related {\em worst-case} results 
mentioned in next section.
They show for example that for each $n$ {\em there exist}
graphs on $n$ nodes 
which may not be relabeled ($\alpha$),
that require in the worst case
$\Omega ( n^3 )$ bits to store
a \term{full information} shortest path routing scheme.
We prove for the same model 
that for {\em almost all} graphs full information routing
$n^3 /4$ bits in total is necessary and sufficient (asymptotically).

\subsection{Related Work}

Previous upper- and lower bounds
on the total number of bits
necessary and sufficient to store the routing scheme
in worst-case static communication
networks
are due to Peleg and Upfal~\cite{PelU89},
and Fraigniaud and Gavoille~\cite{FraG95b}.

In \cite{PelU89}
it was shown that for any stretch factor $s \geq 1$, the total number
of bits required to store the routing scheme for some $n$-node graph is
at least $\Omega ( n^{1+1/(2s+4)} )$ and that there exist
routing schemes for all $n$-node graphs, with stretch factor $s=12k+3$,
using $O(k^3 n^{1+1/k} \log n)$ bits in total.
For example, with stretch factor $s=15$ we have $k=1$
and their method guarantees
$O(n^2\log n)$ bits to store the routing scheme.
The lower bound is shown in the model where nodes may be arbitrarily relabeled
and where nodes know their neighbours (II \mand $\gamma$).
Free port-assignment in conjunction with a model where the neighbours are known
(II) can, however, not be allowed. Otherwise, each node would
gain $n \log n$ bits to store the routing function in (see the footnote to
model II).

Fraigniaud and Gavoille~\cite{FraG95b} showed that
for stretch factors $s<2$
there are routing schemes that require
a total of $\Omega (n^2)$ bits to be stored
in the worst case if nodes may be relabeled by permutation ($\beta$).
This was improved for shortest path routing by 
Gavoille and P{\'{e}}renn\`{e}s~\cite{GavP96} who showed that
for each $d \leq n$
there are shortest path routing schemes that require
a total of $\Omega (n^2 \log d)$ bits to be stored
in the worst case for some graphs with maximal degree $d$,
if nodes may be relabeled by permutation and
the port-assignment may be changed (IB \mand $\beta$).
This last result is clearly optimal for the worst case, both for
general networks ($d= \Theta(n)$) and bounded degree networks ($d < n$).
In \cite{KraK95} it was shown that
 for each $d \geq 3$ there are networks for which
any routing scheme with stretch factor $<2$
requires a total of $\Omega (n^2/\log^2  n)$ bits.

 {\em Interval routing} on a graph $G=(V,E)$, $V=\{1, \ldots ,n\}$
is a routing strategy where for each
node $i$, for each incident edge $e$ of $i$, a
(possibly empty) set of
pairs of node labels represent disjoint intervals with
wrap-around. Each pair indicates the initial edge on a shortest
path from $i$ to any node in the interval, and for each node $j \neq i$
there is such  a pair. We are allowed to permute the labels of
graph $G$ to optimize the interval setting.

\cite{GavP96}
shows that there exist graphs for each bounded degree $d \geq 3$ such that
for each interval routing scheme, each of $\Omega (n)$ edges are
labeled by $\Omega (n)$ intervals.
This shows that interval routing can be worse than straightforward
coding of routing tables which can be trivially done in
$O(n^2 \log d)$ bits total.
(This improves \cite{KraK95} showing
that there exist graphs such that for each interval routing scheme
some incident edge on each of $\Omega (n)$
nodes is labeled by $\Omega (n)$ intervals, and
for each $d \geq 3$ there are graphs of maximal
node degree $d$ such that for each interval routing scheme
  some incident edge on each of $\Omega (n)$
  nodes is labeled by $\Omega (n/\log n)$ intervals.)

The paper \cite{FlaLM95} contains history and background
on the compactness (or lack of it) of interval routing using probabilistic
proof methods.
To the best of our knowledge,
one of the authors of that paper, Jan van Leeuwen, was the first to
formulate explicitly the question of what exactly
is the minimal size of the routing functions, and he recently
drew also our attention to this group
of problems. 

\section{Kolmogorov Complexity}

The Kolmogorov complexity, \cite{Kol65},
of $x$ is the length of the
{\em shortest} effective description of $x$.
That is, the \term{Kolmogorov complexity} $C(x)$ of
a finite string $x$ is simply the length
of the shortest program, say in
FORTRAN (or in Turing machine codes)
encoded in binary, which prints $x$ without any input.
A similar definition holds conditionally, in the sense that
$C(x|y)$ is the length of the shortest binary program
which computes $x$ given $y$ as input. It can be shown that
the Kolmogorov complexity is absolute in the sense
of being independent of the programming language,
up to a fixed additional constant term which depends on the programming
language but not on $x$. We now fix one canonical programming
language once and for all as reference and thereby $C()$.

For the theory and applications, see \cite{LiV93}.
Let $x,y,z \in {\cal N}$, where
${\cal N}$ denotes the natural
numbers. Identify
${\cal N}$ and $\{0,1\}^*$ according to the
correspondence $(0, \epsilon ), (1,0), (2,1)$, $(3,00)$, $(4,01), \ldots$.
Hence, the length $|x|$ of $x$ is the number of bits
in the binary string $x$.
Let $T_1 ,T_2 , \ldots$ be a standard enumeration
of all Turing machines.
Let $\langle \cdot ,\cdot \rangle$ be a standard invertible
effective bijection from ${\cal N} \times {\cal N}$
to ${\cal N}$. This can be iterated to
$\langle  \langle \cdot , \cdot \rangle , \cdot \rangle$.
\begin{definition}
\rm
Let $U$ be an appropriate universal Turing machine
such that $U(\langle \langle i,p \rangle ,y \rangle ) =
T_i (\langle p,y\rangle)$ for all $i$ and $\langle p,y\rangle$.
The \term{Kolmogorov complexity} of $x$ given $y$ (for
free) is
\[C(x|y) = \min\{|p|: U (\langle p,y\rangle)=x , p \in \{0,1\}^*
 \}. \]
\end{definition}

\subsection{Kolmogorov random graphs}

One way to express
irregularity or \term{randomness} of an individual network
topology is by a modern notion of randomness
like Kolmogorov complexity.
A simple counting argument shows that for each
$y$ in the condition and each length $n$
there exists at least one $x$ of length $n$ which
is \term{incompressible} in the sense of $C(x|y) \geq n$,
$50\%$ of all $x$'s of length $n$ is incompressible but for 1 bit
($C(x|y) \geq n-1$), $75\%$ of all $x$'s is incompressible
but for 2 bits ($C(x|y) \geq n-2$) and in general
a fraction of $1 - 1/2^c$ of all strings cannot be compressed
by more than $c$ bits, \cite{LiV93}.
\begin{definition}\label{def.gc}
\rm
Each labeled graph $G=(V,E)$ on $n$ nodes $V=\{1,2,\ldots, n\}$
can be coded
by a binary string $E(G)$ of length $n(n-1)/2$. We enumerate
the $n(n-1)/2$ possible edges $\edge{u}{v}$ in a graph on $n$ nodes
in standard lexicographical order without repetitions
and set the $i$th bit in the string to 1 if the $i$-th
edge is present and to 0 otherwise. Conversely, each
binary string of length $n(n-1)/2$ encodes a graph on $n$
nodes. Hence we can identify each such graph with
its corresponding binary string. 
\end{definition}
We define the high complexity
graphs in a particular family ${\cal G}$ of graphs.
\begin{definition}\label{def.rg}
\rm
A labeled graph $G$ on $n$
nodes of a family ${\cal G}$ of graphs  
has {\em randomness deficiency}\index{randomness deficiency}
at most
$\delta (n)$ and is called
$\delta (n)$-{\em random} in ${\cal G}$
if it satisfies
\begin{equation}\label{eq.KG}
 C(E(G)|n, \delta , {\cal G}) \geq \log |{\cal G}| - \delta (n).
\end{equation}
In this paper we use ${\cal G}$ is the set of all labeled  graphs on $n$ 
nodes. Then, $\log |{\cal G}| = n(n-1)/2$, that is, precisely the length
of the encoding of Definition~\ref{def.gc}.
In the sequel we say just 
`$\delta (n)$-{\em random}' with ${\cal G}$ understood.
\end{definition}
Elementary counting shows
that {\em a fraction} of at least
\[ 1 - 1/2^{\delta (n)} \]
of all labeled graphs on $n$ nodes in ${\cal G}$ has that high complexity,
\cite{LiV93}. 

\subsection{Self-Delimiting Binary Strings}
We need the notion of self-delimiting binary
strings.
\begin{definition}
\rm
We call $x$ a %
\it prefix %
\rm of $y$ if there is a $z$ such that
$y = xz$. A set $A \subseteq   \{ 0,1 \}^*$ is
\term{prefix-free}, if no element in $A$ is the prefix of another
element in $A$. A 1:1 function
$E: \{ 0, 1 \}^*  \rightarrow  \{0,1\}^*$
(equivalently, $E: {\cal N}  \rightarrow  \{0,1\}^*$)
defines a \term{prefix-code}
if its range is prefix-free.
A simple prefix-code we use throughout is obtained by reserving one
symbol, say 0, as a stop sign and encoding
\begin{eqnarray*}
\bar x & = & 1^{|x|} 0 x, \\
| \bar x | & = & 2|x| + 1.
\end{eqnarray*}
Sometimes we need the shorter prefix-code $x'$:
\begin{eqnarray*}
x' & = & \overline{|x|} x, \\
|x' | & = & |x| + 2 \ceil{\log (|x|+1)} + 1.
\end{eqnarray*}
We call $\bar x$ or $x'$ a \term{self-delimiting} version
of the binary string $x$.
We can effectively recover both $x$ and $y$ unambiguously
from the binary strings $\bar x y$ or $x'y$.
For example, if $\bar x y = 111011011$, then $x = 110$ and $y = 11$.
If $ \bar x  \bar y = 1110110101$ then $x = 110$ and $y = 1$.
The self-delimiting form $x' \ldots y'z$
allows the concatenated binary sub-descriptions to
be parsed and unpacked into the individual items $x, \ldots, y,z$;
the code $x'$ encodes a separation delimiter for $x$,
using
$2 \ceil{\log (|x|+1)}$ extra bits, and so on, \cite{LiV93}.
\end{definition}
\subsection{Topological Properties of Kolmogorov Random Graphs}
High complexity labeled graphs have many specific topological
properties which seems to contradict their randomness.
However, randomness is not `lawlessness' but rather
enforces strict statistical regularities. For example,
to have diameter exactly two. Note that randomly generated graphs
have diameter two with high probability. In another paper
\cite{BLV96} two of us explored the relation between high probability properties
of random graphs and properties of individual Kolmogorov random graphs.
For this discussion it is relevant to mention that,
in a precisely quantified way, {\em every}
Kolmogorov random graph individually possesses all simple properties
which hold with high probability for randomly generated graphs.

\begin{lemma}\label{lem.degree}
The degree $d$ of every node of a $\delta (n)$-random  
labeled graph on $n$ nodes satisfies
$$|d-(n-1)/2| = O \left( \sqrt{(\delta(n) + \log n) n} \right).$$
\end{lemma}
\begin{proof}
Assume that there is a node such
that the deviation of its degree $d$
from $(n-1)/2$ is greater than $k$, that is,
$|d-(n-1)/2| > k$. From the
lower bound on $C(E(G)|n, \delta, {\cal G})$ corresponding to the assumption
that $G$ is random in ${\cal G}$, we can estimate an upper bound
on $k$, as follows:

In a description of $G=(V,E)$ given $n, \delta$  we can
indicate which edges
are incident on node $i$ by giving the index
of the interconnection pattern (the characteristic sequence
of the set $V_i = \{j \in V - \{i\}: (i,j) \in E \}$  in $n-1$ bits where
the $j$th bit is 1 if $j \in V_i$ and 0 otherwise)  in the ensemble
of
\begin{equation}\label{eq.m}
m= \sum_{|d-(n-1)/2| > k} {{n-1} \choose d}
\leq 2^n e^{-2k^2/3(n-1)}
\end{equation}
possibilities.
The last inequality follows from
a general estimate of the tail probability of the binomial distribution,
with $s_n$ the number of successful outcomes in $n$ experiments
with probability of success $p = \frac{1}{2}$.
Namely, by Chernoff's bounds\index{Chernoff bounds},
in the form used in \cite{AV79,LiV93},
\begin{equation}\label{chernoffzerotex1}
\Pr (|s_n -pn| > k ) \leq 2e^{-k^2 /3pn } .
\end{equation}
To describe $G$ it then suffices
to modify the old code of $G$ by
prefixing it with
\begin{romannum}
\item
A description of this discussion in $O(1)$ bits;
\item
the identity of
node $i$ in $\lceil \log (n+1) \rceil$ bits;
\item
the value of $d$ in $\lceil \log (n+1) \rceil$ bits,
possibly adding nonsignificant 0's to pad up to this amount;
\item
the index of the interconnection pattern in $ \log m $ bits
(we know $n,k$ and hence $\log m$);
followed by
\item
the old code for $G$
with the bits in the code denoting the presence
or absence of the possible edges that are incident
on node $i$ deleted.
\end{romannum}
Clearly, given $n$ we can reconstruct
the graph $G$ from the new description. The total description we have
achieved is an effective program  of
\[\log m + \log nk + n(n-1)/2 - n +O(1)\]
bits. This must be at least the length of the shortest
effective binary program, which is
$C(E(G)|n, \delta, {\cal G} )$, satisfying Equation~\ref{eq.KG}.
Therefore,
\[ \log  m  \geq n - \log nk - O(1) - \delta(n). \]
Since we have estimated in Equation~\ref{eq.m} that
\[ \log  m \leq n - (2k^2/3(n-1)) \log e , \]
it follows that
$k \leq \sqrt{\frac{3}{2}(\delta (n) + \log nk + O(1)) (n-1)/\log e} $.
\qquad
\end{proof}

\begin{lemma}\label{lem.diam}
Every $o(n)$-random labeled graph on $n$ nodes 
has diameter 2.
\end{lemma}
\begin{proof}
The only graphs with diameter 1 are the complete graphs which
can be described in $O(1)$ bits, given $n$, and hence are not random.
It remains to consider $G=(V,E)$ is an $o(n)$-random graph with diameter
greater than 2. 
which contradicts Eq.~(\ref{eq.KG}) from some $n$ onwards.

It remains to consider $G=(V,E)$ is an $o(n)$-random graph with diameter
greater than 2. Let $i,j$  be a pair of nodes with distance
greater than 2. Then we can describe $G$ by modifying the old
code for $G$ as follows:
\begin{romannum}
\item
A description of this discussion in $O(1)$ bits;
\item
The identities of $i < j$ in  $2  \log  n$ bits;
\item
The old code $E(G)$ of $G$ with all bits representing
presence or absence of an edge $(j,k)$ between $j$
and each $k$ with $(i,k) \in E$
deleted. We know that all the bits representing such edges
must be 0 since the existence of any such edge shows that $(i,k),(k,j)$
is a path of length 2 between $i$ and $j$, contradicting the assumption
that $i$ and $j$ have distance $>2$.  This way we save at least $n/4$
bits, since we save bits for as many edges $(j,k)$ as there
are edges $(i,k)$, that is, the degree of $i$, which is $n/2 \pm o(n)$
by Lemma~\ref{lem.degree}.
\end{romannum}
Since we know the identities of
$i$ and $j$ and the nodes adjacent to $i$ (they are
in the prefix of code $E(G)$ where no bits have been
deleted),  we can reconstruct $G$ from this discussion and the new description, given $n$. Since by
Lemma~\ref{lem.degree} the degree of $i$ is at least $n/4$,
the new description of $G$, given $n$,
requires at most
\[ n(n-1)/2  - n/4 + O(\log n) \]
bits, which contradicts Equation~\ref{eq.KG} 
for large $n$.
\qquad
\end{proof}

\begin{lemma}\label{lem.logn_edges}
Let $c \geq 0$ be a fixed constant and let
$G$ is a $c \log n$-random labeled graph.
Then from each node $i$ all other nodes
are either directly connected to $i$
or are directly connected to one of the least
$(c+3) \log n $ nodes directly adjacent to $i$.
\end{lemma}
\begin{proof}
Given $i$, let $A$ be
the set of the least $(c+3) \log n$ nodes directly adjacent to $i$.
Assume by way of contradiction that
 there is a node $k$ of $G$ that is not directly connected
to a node in $A \bigcup \{i\}$.
We can describe $G$ as follows:
\begin{romannum}
\item
A description of this discussion in $O(1)$ bits;
\item
A literal description of $i$ in $ \log n$ bits;
\item
A literal description of the presence or absence of edges between
$i$ and the other nodes in $n-1$ bits;
\item
A literal description of $k$ and its incident edges
in $\log n + n-2 -  (c+3) \log n$ bits;
\item
The encoding $E(G)$ with the edges
incident with nodes $i$ and $k$ deleted, saving at least $2n- 2$ bits.
\end{romannum}
Altogether the resultant description has
\[n(n-1)/2 + 2 \log n  + 2n-3 - (c+3)\log n
- 2n + 2 \]
bits, which contradicts the $c \log n$-randomness of $G$
by Equation~\ref{eq.KG}.
\qquad
\end{proof}

In the description we have
explicitly added the adjacency pattern of node $i$, which
we deleted later again. This zero-sum swap is necessary
to be able to unambiguously identify the adjacency pattern of $i$
in order to reconstruct $G$.
Since we know the identities of
$i$ and the nodes adjacent to $i$ (they are the
prefix where no bits have been
deleted),  we can reconstruct $G$
from this discussion and the new description, given $n$.

\section{Upper Bounds}

We give methods to route messages over Kolmogorov random
graphs with compact routing schemes. Specifically we
show that in general (on almost all graphs) one can use shortest path
routing schemes occupying at most $O(n^2)$ bits.
If one can relabel the graph in advance, and if nodes know their
neighbours, shortest path routing schemes are shown to occupy only
$O(n \log^2 n)$
bits. Allowing stretch factors larger than one reduces the
space requirements---to $O(n)$ bits for stretch factors
of $O(\log n)$.

Let $G$ be an $O(\log n)$-random labeled graph on $n$ nodes.
By Lemma~\ref{lem.logn_edges} we know that from each
node $u$ we can shortest path route to each node $v$
through the least $O(\log n)$ directly
adjacent nodes of $u$. By Lemma~\ref{lem.diam},
$G$ has diameter 2. Once
the message reached node $v$ its
destination is either node $v$ or a direct
neighbor of node $v$ (which is known in node $v$ by assumption).
Therefore, routing functions of size
$O(n\log\log n)$ bits per node can be used to
do shortest path routing. However, we
can do better.
 
\begin{theorem}\label{theo.ub1}
Let $G$ be an $O(\log n)$-random labeled graph on $n$ nodes.
Assume that the
port assignment may be changed or nodes know their neighbours
(IB \mor II). Then, for shortest path routing
it suffices to have local routing functions stored in $3n$ bits per node.
Hence the complete routing scheme is represented by $3n^2$ bits.
\end{theorem}
\begin{proof}
Let $G$ be as in the statement of the theorem.
By Lemma~\ref{lem.logn_edges} we know that from each
node $u$ we can route via shortest paths to each node $v$
through the $O(\log n)$ directly
adjacent nodes of $u$ that have the least indexes.
By Lemma~\ref{lem.diam},
$G$ has diameter 2. Once
the message has reached node $v$ its
destination is either node $v$ or a direct
neighbor of node $v$ (which is known in node $v$ by assumption).
Therefore, routing functions of size
$O(n\log\log n)$ can be used to
do shortest-path routing. We
can do better than this.

Let $A_0 \subseteq V$ be the set of nodes in $G$ which are not directly
connected to $u$. Let $v_1 , \ldots , v_m$ be
the $O(\log n)$ least indexed nodes directly
adjacent to node $u$ (Lemma~\ref{lem.logn_edges}) 
through which we can shortest path route to
all nodes in $A_0$.
For $t=1,2 \ldots, l$ define
$A_t =\{w \in A_0 - \bigcup_{s=1}^{t-1} A_s: (v_t,w) \in E\}$.
Let $m_0 = |A_0|$ and define $m_{t+1} = m_t - |A_{t+1}|$.
Let $l$ be the first $t$ such that $m_t < n/\log \log n$.
Then we claim that $v_t$ is connected by an edge in $E$ to at least $1/3$ of
the nodes not connected by edges in $E$ to nodes $u,v_1, \ldots, v_{t-1}$.
\begin{claim}\label{claim.rout1}
$|A_t| > m_{t-1}/3$ for $1 \leq t \leq l$.
\end{claim}
\begin{proof}
Suppose, by way of contradiction,
that there exists a least $t \leq l$  such that
$||A_t| - m_{t-1}/2| \geq m_{t-1}/6$.
Then we can describe $G$, given $n$, as follows.
\begin{romannum}
\item
This discussion in $O(1)$ bits;
\item
Nodes $u,v_t$ in $2 \log n$ bits, padded with 0's if need be;
\item
The presence or absence of edges incident with nodes $u,v_1, \ldots ,v_{t-1}$
in $r= n-1 + \cdots + n - (t-1)$ bits. This gives us
the characteristic sequences of $A_0, \ldots ,A_{t-1}$ in $V$,
where a {\em characteristic sequence} of $A$ in $ V$ is a string of
$|V|$ bits with, for each $v \in V$, the $v$th bit equals $1$ if $v
\in A$ and the $v$th bit is $0$ otherwise;
\item
A self-delimiting description of the characteristic sequence
of $A_t$ in $ A_0- \bigcup_{s=1}^{t-1} A_s$, using
Chernoff's bound\index{Chernoff bounds}
Equation~\ref{chernoffzerotex1},
in at most $m_{t-1} - \frac{2}{3} \left( \frac{1}{6} \right)^2
m_{t-1} \log e + O(\log m_{t-1})$ bits;
\item
The description $E(G)$ with all bits corresponding to the presence or absence
of edges between $v_t$ and the nodes in $A_0-\bigcup_{s=1}^{t-1} A_s$ deleted,
saving $ m_{t-1}$ bits.
Furthermore, we delete also all bits corresponding to
presence or absence of edges incident with $u,v_1,\ldots,v_{t-1}$ saving a
further $r$ bits.
\end{romannum}
This description of $G$ uses at most
\[ n(n-1)/2 + O( \log n) + m_{t-1}  - \frac{2}{3} \left( \frac{1}{6} \right)^2
m_{t-1} \log e
 - m_{t-1} \]
bits, which contradicts the $O(\log n)$-randomness
of $G$ by Equation~\ref{eq.KG},
because $m_{t-1} > n/\log \log n$.
\qquad
\end{proof}
 
Recall that $l$ is the least integer such that
$m_l < n/\log \log n$. We construct the local routing function $F(u)$
as follows.
\begin{romannum}
\item A table of intermediate routing node entries
for all the nodes in $A_0$ in increasing order. For each node $w$
in $\bigcup_{s=1}^l A_s$ we enter in the $w$th position in the table
the unary representation of the least
intermediate node $v$, with $(u,v) , (v,w) \in E$,
followed by a $0$.  For the nodes that are not in
$\bigcup_{s=1}^l A_s$ we enter a $0$ in their position
in the table indicating that an entry for this node can
be found in the second table.
By Claim~\ref{claim.rout1}, the size of this table is bounded by:
\[
n+\sum_{s=1}^{l} \frac{1}{3}
\left( \frac{2}{3} \right)^{s-1} sn \leq
n+ \sum_{s=1}^{\infty} \frac{1}{3}
\left( \frac{2}{3} \right)^{s-1} sn \leq 4n ;
\]
\item A table with explicitly binary coded
intermediate nodes on a shortest path
for the ordered set of the remaining destination nodes. Those nodes
had a $0$ entry in the first table and there are  at most
$m_l < n/ \log \log n$ of them, namely the nodes in
$A_0 -\bigcup_{s=1}^l A_s$. Each entry consists of the code of length
$\log \log n + O(1)$ for the position in increasing order of
a node out of
$v_1 , \ldots , v_m$ with $m = O(\log n)$ by Lemma~\ref{lem.logn_edges}.
Hence this second table requires at most $2n$ bits.
\end{romannum}
The routing algorithm is as follows. The direct neighbors of $u$
are known in node $u$ and are routed without
routing table. If we route from start node $u$
to target node $w$ which is not directly adjacent to $u$,
then we do the following. If node $w$ has an entry in the first table then
route over the edge coded in unary, otherwise find an entry for node $w$ in the
second table.

Altogether, we have $|F(u)| \leq  6n$. Adding another $n-1$ in case the
port assignment may be chosen arbitrarily, this proves the theorem
with $7n$ instead of $6n$.
Slightly more precise counting and choosing $l$ such that $m_l$
is the first such quantity $< n/ \log n$ shows $|F(u)| \leq 3n$.
\qquad
\end{proof}

If we allow arbitrary labels for the nodes, then shortest path routing schemes
of $O(n \log^2 n)$ bits suffice on Kolmogorov random graphs, as witnessed by
the following theorem.
\begin{theorem}\label{theo.ub5}
Let $c\geq 0$ be a constant and let $G$ be a $c \log n$-random labeled  gra\-ph
on $n$ nodes.
Assume that nodes know their
neighbours and nodes may be arbitrarily relabeled (II \mand $\gamma$), and
we allow the use of
labels of
$(1+(c+3)\log n)\log n$ bits. Then we can
shortest path route with local routing functions stored in
$O(1)$ bits per node (hence the complete routing scheme is represented by
$(c+3)n \log^2 n + n \log n + O(n)$ bits).
\end{theorem}
\begin{proof}
Let $c$ and $G$ be as in the statement of the theorem.
By Lemma~\ref{lem.logn_edges} we know that from each
node $u$ we can shortest path route to each node $w$
through the first $(c+3) \log n$ directly adjacent
nodes $f(u) = v_1, \ldots , v_m$ of $u$.
By lemma~\ref{lem.diam}, $G$ has diameter 2.
Relabel $G$ such that the label
of node $u$ equals $u$ followed by the original labels of the first
$(c+3) \log n$
directly adjacent nodes $f(u)$. This new label occupies
$(1 + (c+3)\log n) \log n$ bits. To route from source $u$ to destination $v$
do the following.

If $v$ is directly adjacent to $u$ we route to $v$
in $1$ step in our model (nodes know their neighbours). If $v$ is not directly
adjacent to $u$, we consider
the immediate neighbours $f(v)$ contained in the name of $v$. By
Lemma~\ref{lem.logn_edges} at least one of the neighbours of $u$ must have
a label whose original label (stored in the first $\log n$ bits
of its new label) corresponds
to one of the labels in $f(v)$. Node $u$ routes the message to any such
neighbour. This routing function can be stored in $O(1)$ bits.
\qquad
\end{proof}

Without relabeling routing using less than $O(n^2)$ bits is possible if we
allow stretch factors larger than $1$. The next three theorems clearly show a
trade-off between the stretch factor and the size of the routing scheme.
\begin{theorem}\label{theo.ub4}
Let $c \geq 0$ be a constant and let $G$ be a  $c \log n$-random
labeled graph on $n$ nodes. Assume that  nodes know their neighbours (II).
For routing with any stretch factor $>1$ 
it suffices to have
$n - 1 - (c+3) \log n$ nodes with local routing functions
stored in at most $\ceil{\log (n+1)}$ bits per node,
and $1 + (c+3)\log n $ nodes with local routing functions stored in $3n$ bits
per node (hence the complete routing scheme is represented by less
than $(3c+20) n  \log n$ bits). Moreover, the stretch is at most 1.5.
\end{theorem}
\begin{proof}
Let $c$ and $G$ be as in the statement of the theorem.
By Lemma~\ref{lem.logn_edges} we know that from each
node $u$ we can shortest path route to each node $w$
through the first $(c+3) \log n$ directly adjacent
nodes $v_1, \ldots , v_m$ of $u$. By Lemma~\ref{lem.diam},
$G$ has diameter 2. Consequently, each node in $V$
is directly adjacent to some node in $B=\{u,v_1, \ldots , v_m\}$.
Hence, it suffices to select the nodes of $B$
as routing centers and store, in each node $w \in B$, a shortest path
routing function $F(w)$ to all other nodes, occupying
$3n$ bits (the same routing function as constructed
in the proof of Theorem~\ref{theo.ub1} if the neighbours are known).
Nodes $v \in V-B$ route any destination unequal to their own label to
some fixed directly adjacent node $w \in B$. Then
$|F(v)| \leq \ceil{\log (n+1)} + O(1)$, and this gives the
bit count in the theorem

To route from a originating node $v$ to a target node $w$ the following
steps are taken.  If $w$ is directly adjacent to $v$ we route to $w$
in $1$ step in our model. If $w$ is not directly adjacent to $v$
then we first route in $1$ step from $v$ to its directly connected
node in $B$, and then via a shortest path to $w$. Altogether, this
takes either 2 or 3 steps whereas the shortest path has length 2.
Hence the stretch factor is at most $1.5$ which for graphs
of diameter 2 (\ie all $c \log n$-random graphs by Lemma~\ref{lem.diam}) is the
only possibility between stretch factors 1 and 2.
This proves the theorem.
\qquad
\end{proof}

\begin{theorem}\label{theo.ub2}
Let $c \geq 0$ be a constant and let $G$ be a  $c \log n$-random
labeled graph on $n$ nodes. Assume that the nodes know their neighbours (II).
For routing with stretch factor $2$
it suffices to have $n-1$ nodes with local routing functions stored
in at most $\log \log n $ bits per node
and $1$ node with its local routing function stored in $3n$ bits
(hence the complete routing scheme is represented by
$n \log \log n + 3n$ bits).
\end{theorem}
\begin{proof}
Let $c$ and $G$ be as in the statement of the theorem.
By Lemma~\ref{lem.diam}, $G$ has diameter 2. Therefore the following routing
scheme has stretch factor $2$. Let node $1$ store a shortest path routing
function. All other nodes only store a shortest path to node $1$. To route
from a originating node $v$ to a target node $w$ the
following steps are taken. If $w$ is an immediate neighbour of $v$, we route
to $w$ in $1$ step in our model. If not, we first route the message to node $1$
in at most $2$ steps, and then from node $1$ through a node $v$ to node $w$ in
again $2$ steps. Because node $1$ stores a shortest path routing function,
either
$v=w$ or $w$ is a direct neighbour of $v$.

Node $1$ can store a shortest path routing function in at most $3n$ bits using
the same construction as used in the proof of Theorem~\ref{theo.ub1} (if the
neighbours are known). The
immediate neighbours of $1$ either route to $1$ or directly to the destination
of the message. For these nodes, the routing function occupies $O(1)$ bits.
For nodes $v$ at distance $2$ of node $1$ we use Lemma~\ref{lem.logn_edges},
which tells us that we can shortest path route to node $1$
through the first $(c+3) \log n$ directly adjacent nodes of $v$. Hence, to
represent this edge takes $\log \log n + \log(c+3)$ bits
and hence the local routing function $F(v)$ occupies at most
$\log \log n +O(1)$ bits.
\qquad
\end{proof}

\begin{theorem}\label{theo.ub3}
Let $c\geq0$ be a constant and let $G$ be a $c \log n$-random labeled
graph on $n$ nodes.
Assume that nodes know their neighbours (II).
For routing with stretch factor $(c+3)\log n$ 
it suffices to have
local routing functions stored in $O(1)$ bits per node
(hence the complete routing scheme is represented by $O(n)$ bits).
\end{theorem}
\begin{proof}
Let $c$ and $G$ be as in the statement of the theorem.
From Lemma~\ref{lem.logn_edges} we know that from each
node $u$ we can shortest path route to each node $v$
through the first $(c+3) \log n$ directly adjacent nodes of $u$.
By Lemma~\ref{lem.diam},
$G$ has diameter 2. So the local routing function --- representable in $O(1)$
bits --- is to route directly
to the target node if it is a directly adjacent node,
otherwise to simply traverse the first
$(c+3) \log n$ incident edges of the starting node and look in
each of the visited nodes whether the target node is a directly
adjacent node. If so, the message is forwarded to that node, otherwise it is
returned to the starting node for trying the next node.
Hence each message for a destination at distance $2$ traverses at
most $2 (c+3) \log n$ edges.

Strictly speaking we do not use routing tables at all.
We use the fact that a message can go back and forth 
several times to a node. The header of the message can code some extra
information as a tag ``failed.'' In this case it is possible to describe
an $O(1)$ bit size routing function allowing to extract the header from the destination
without knowing about $\log n$, for example by the use of self-delimiting
encoding.
\qquad
\end{proof}

\begin{theorem}\label{theo.ub6}
Let $G$ be an $O( \log n)$-random labeled gra\-ph on $n$ nodes. Assume that
nodes do not know their neighbors and
relabeling and changing the port assignment is not allowed
(IA \mand $\alpha$). 
Then, for shortest path routing it
suffices that each local routing function uses  $(n/2) \log n(1+o(1))$
bits (hence the complete routing scheme 
uses at most $(n^2/2) \log n (1+o(1))$
bits to be stored). 
\end{theorem}
\begin{proof}
At each node we can give the neighbors by the positions of the 1's in
a binary string of length $n-1$.
 Since each node has at most
 $n/2 + o(n)$ neighbours by Lemma~\ref{lem.degree},
a permutation of port-assignments to neighbors
 can have Kolmogorov complexity at most
$(n/2) \log n (1+o(1))$~\cite{LiV93}. This permutation $\pi$ describes 
part of the local routing function by for each direct neighbour 
determining the port to route
messages for that neighbour over. If $G$ is $O( \log n)$-random
then we only require $O(n)$ bits additional routing
information in each node by Theorem~\ref{theo.ub1}. Namely, because
the assignment of ports (outgoing edges) to direct neighbors
is known by permutation $\pi$ we can use
an additional routing table in $3n$ bits per node to route
to the remaining non-neighbor nodes as described
in the proof of Theorem~\ref{theo.ub1}. In total this gives
$(n^2/2) \log n (1+ o(1))$ bits. 
\qquad
\end{proof}

Our last theorem of this section determines the upper bounds
for full information
shortest path routing schemes on Kolmogorov random graphs.
\begin{theorem}\label{th.fullinfoub}
For full-information shortest path routing on
$o(n)$-random labeled gra\-phs on $n$ nodes
where relabeling is not allowed ($\alpha$),
the local routing function
occupies at most $n^2/4 + o(n^2)$ bits for every node
(hence the complete routing scheme takes at most $n^3/4 + o(n^3)$
bits to be stored).
\end{theorem}
\begin{proof}
	Since for $o(n)$-random labeled graphs on $n$ 
	the node degree of every node is $n/2 + o(n)$ by
	Lemma~\ref{lem.degree}, we can
	in each source node describe
	the appropriate outgoing edges (ports) for each destination node
	by the 1's in a binary string of length $n/2 + o(n)$.
	For each source node it suffices to store at
	most $n/2 + o(n)$ such binary strings
	corresponding to the non-neighboring destination nodes.
	In each node we can give the neighbors by the positions of the 1's
	in a binary string of length $n-1$.
	Moreover, in each node we can give the permutation of port assignments
	to neighbors in $(n/2) \log n (1+o(1))$ bits.  
	This leads to a total of at most $(n^2/4)(1+o(1))$ bits
	per node and hence to $(n^3/4) (1+ o(1))$ bits to store the overall
	routing scheme.
\qquad
\end{proof}

\section{Lower Bounds}

The first two theorems of this section together show that indeed $\Omega(n^2)$
bits are necessary to route on Kolmogorov random graphs in all models we
consider, except for the models where nodes know their neighbours {\em and}
label permutation or relabeling is allowed 
(II \mand $\beta$ or II \mand $\gamma$).
Hence the upper bound in Theorem~\ref{theo.ub1} is tight up to order of
magnitude.

\begin{theorem}\label{theo.lb1}
For shortest path routing in $o(n)$-random labeled
 graphs where relabeling is not
allowed and nodes know their neighbours (II \mand $\alpha$), each local routing
function must be stored in at least $n/2 - o(n)$ bits per node
(hence the complete routing scheme requires at least $n^2/2 - o(n^2)$ bits
to be stored).
\end{theorem}
\begin{proof}
Let $G$ be an $o(n)$-random graph. Let $F(u)$
be the local routing function of node $u$ of $G$, and let $|F(u)|$
be the number of bits used to store $F(u)$. Let $E(G)$ be the standard
encoding of $G$ in $n(n-1)/2$ bits as in Definition~\ref{def.gc}.
We now give another way to describe $G$ using some
local routing function $F(u)$.
\begin{romannum}
\item
A description of this discussion in $O(1)$ bits;
\item
A description of $u$ in exactly $\log n$ bits, padded with 0's if needed;
\item
A description of the presence or absence of edges between $u$ and the
other nodes in $V$ in $n-1$ bits;
\item
A self-delimiting description of $F(u)$ in $|F(u)| + 2\log |F(u)|$ bits
\item
The code $E(G)$ with all bits deleted
corresponding to edges $(v,w) \in E$ for
each $v$ and $w$ such that $F(u)$ routes messages to $w$ through
the least intermediary node $v$. This saves at least $n/2 - o(n)$ bits
since there are at least $n/2-o(n)$ nodes $w$ such that
$(u,w) \notin E$ by Lemma~\ref{lem.degree}, and since the diameter
of $G$ is 2 by Lemma~\ref{lem.diam} there is a shortest
path $(u,v), (v,w) \in E^2$ for some $v$.
Furthermore, we delete all bits corresponding to the presence or absence
of edges between $u$ and the other nodes in $V$, saving another $n-1$
bits. This corresponds to the $n-1$ bits for edges connected to $u$
which we added in one connected block above.
\end{romannum}
In the description we have
explicitly added the adjacency pattern of node $u$ which
explicitly added the adjacency pattern of node $u$ which
we deleted elswewhere. This zero-sum swap is necessary
to be able to unambiguously identify the adjacency pattern of $u$
in order to reconstruct $G$ given $n$, as follows:
Reconstruct the bits corresponding to the deleted edges
using $u$ and $F(u)$
and subsequently insert them in the
appropriate positions of the remnants of $E(G)$. We can do so
because these positions can be simply reconstructed in increasing order.
In total this new description has
\[ n(n-1)/2 + O(1)+ O(\log n) + |F(u)|
- n/2 + o(n) \]
which must be at least $n(n-1)/2 -o(n)$ by Equation~\ref{eq.KG}.
Hence, $|F(u)| \geq n/2 - o(n)$, which proves the theorem.
\qquad
\end{proof}

\begin{theorem}\label{theo.lb1-a}
Let $G$ be an $o(n)$-random labeled gra\-ph on $n$ nodes. Assume that
the neighbours are not known (IA \mor IB) but relabeling is allowed
($\gamma$).
Then, for shortest path routing
the complete routing scheme requires at least $n^2/32 - o(n^2)$ bits
to be stored.
\end{theorem}
\begin{proof}
In the proof of this theorem we need the following combinatorial result.
\begin{claim}\label{cl.dopjes-sub}
Let $k$ and $n$ be arbitrary natural numbers such that $1 \le k \le n$.
Let $x_i$, for $1 \le i \le k$, be natural numbers such that $x_i \ge 1$.
If $\sum_{i=1}^k x_i = n$, then
\[
\sum_{i=1}^k \ceil{\log x_i} \le n - k
\]
\end{claim}
\begin{proof}
By induction on $k$. If $k=1$, then $x_1 = n$ and clearly
$\ceil{\log n} \le n-1$ if $n \ge 1$. Supposing the claim holds for
$k$ and arbitrary $n$ and $x_i$, we now prove it for $k' = k+1$, $n$
and arbitrary $x_i$. Let $\sum_{i=1}^{k'} x_i = n$. Then
$\sum_{i=1}^{k} x_i = n - x_{k'}$. Now
\[
\sum_{i=1}^{k'} \ceil{\log x_i} = \sum_{i=1}^k \ceil{\log x_i} +
	\ceil{\log x_{k'}}
\]
By the induction hypothesis the first term on the right-hand side
is less than or equal to $n - x_{k'} - k$, so
\[
\sum_{i=1}^{k'} \ceil{\log x_i} \le
	n - x_{k'} - k + \ceil{\log x_{k'}}
	= n - k' + \ceil{\log x_{k'}} + 1 - x_{k'}
\]
Clearly $ \ceil{\log x_{k'}} + 1 \le x_{k'} $ if $x_{k'} \ge 1$, which
proves the claim.
\qquad
\end{proof}

Recall that in model $\gamma$ each router must be able to output its own
label. Using the routing scheme we can enumerate the labels of all nodes.
If we cannot enumerate the labels of all nodes using less than $n^2/32$ bits
of information, then the routing scheme requires at least that many
bits of information and we are done. 
So assume we can (this includes models $\alpha$ and $\beta$
where the labels are not charged for, but can be described using $\log n$
bits). Let $G$ be an
$o(n)$-random graph.
\begin{claim}\label{cl.dopjes}
\rm
Given the labels of all nodes, we can describe the interconnection pattern
of a node $u$ using the local routing function of node $u$ plus an additional
$n/2 + o(n)$ bits.
\end{claim}
\begin{proof}
Apply the local routing function to each of the labels of the nodes in
turn (these are given by assumption). This will return for each edge a list of
destinations reached over that
edge. To describe the interconnection pattern it remains to encode, for each
edge, which of the destinations reached is actually its immediate neighbour.
If edge $i$ routes $x_i$ destinations, this will cost $\ceil{\log x_i}$
bits.
By Lemma~\ref{lem.degree} the degree of a node in $G$ is at least $n/2 - o(n)$.
Then in total, $\sum_{i=1}^{n/2 - o(n)} \ceil{\log x_i} $
bits will be sufficient; separations need not be encoded because they can be
determined using the knowledge of all $x_i$'s. Using Claim~\ref{cl.dopjes-sub}
finishes the proof.
\qquad
\end{proof}

Now we show that there are $n/2$ nodes in $G$ whose local routing function
requires at least $n/8 - 3 \log n$ bits to describe (which implies the
theorem).

Assume, by way of contradiction, that there are $n/2$ nodes in $G$ whose
local routing function requires at most $n/8 - 3 \log n$ bits to describe.
Then we can describe $G$ as follows:
\begin{romannum}
\item A description of this discussion in $O(1)$ bits,
\item The enumeration of all labels in at most $n^2/32$ (by assumption),
\item A description of the $n/2$ nodes in this enumeration in at most $n$ bits,
\item The interconnection patterns of these $n/2$ nodes in
  $n/8 - 3 \log n$ plus $n/2 + o(n)$ bits each
  (by assumption, and using Claim~\ref{cl.dopjes}).
  This amounts to $n/2 (5n/8 - 3 \log n) + o(n^2)$ bits in total, with
  separations encoded in another $n \log n$ bits,
\item The interconnection patterns of the remaining $n/2$ no\-des
  {\em only among
  themselves\/} using the standard encoding, in $1/2 (n/2)^2$ bits.
\end{romannum}
This description altogether uses
\begin{eqnarray*}
\lefteqn{O(1) + n^2/32 + n + n/2 (5n/8 - 3 \log n) + } \\
&&      + o(n^2) + n \log n + 1/2 (n/2)^2 =  \\
&&	= n^2/2 - n^2/32 + n + o(n^2) - n/2 \log n
\end{eqnarray*}
bits, contradicting the $o(n)$-randomness of $G$ by Eq.~(\ref{eq.KG}).
We conclude that on at least $n/2$ nodes a total of $n^2/16 - o(n^2)$ bits are
used to store the routing scheme.
\qquad
\end{proof}

If neither relabeling nor changing the port assignment is allowed, the next
theorem implies that for shortest path routing on
almost all such `static' graphs one
cannot do better than storing part of
the routing tables literally, in $(n^2 /2)  \log n$
bits. Note that it is known \cite{GavP96} 
that there are {\em worst-case} graphs 
(even in models where relabeling
is allowed) such that $n^2 \log n - O(n^2)$ bits are required 
to store the routing scheme, and this matches the trivial upper
bound for all graphs exactly. But in our Theorem~\ref{theo.lb2}
we show that in a certain restricted model for {\em almost all} graphs
asymptotically $(n^2 /2) \log n$ bits are required and 
by Theorem~\ref{theo.ub6} that many bits are also sufficient.

\begin{theorem}\label{theo.lb2}
Let $G$ be an $o(n)$-random labeled gra\-ph on $n$ nodes. Assume that
nodes do not know their neighbors and
relabeling and changing the port assignment is not allowed
(IA \mand $\alpha$). Then, for shortest path routing 
each local routing function must be stored in
at least $(n/2) \log n - O(n)$ bits per node
(hence the complete routing scheme requires at least $(n^2/2) \log n - O(n^2)$
bits to be stored).
\end{theorem}
\begin{proof}
If the graph cannot
be relabeled and the port-assignment cannot be changed, the adversary can set
the port-assignment of each node to correspond to a permutation of the
destination nodes. Since each node has at least
 $n/2 - o(n)$ neighbours by Lemma~\ref{lem.degree},
such a permutation can have Kolmogorov complexity as high as
$(n/2) \log n - O(n)$~\cite{LiV93}. Because the neighbours are not known,
the local routing function must for each neighbor node determine 
the port to route
messages for that neighbor node over. Hence the local routing function
completely describes the permutation, given the neighbors, 
and thus it must occupy at least
$(n/2) \log n - O(n)$ bits per node. 
\qquad
\end{proof}

Note that in this model (IA \mand $\alpha$) we can trivially find
by the same method a lower bound of $n^2 \log n - O(n^2)$ bits
for specific graphs like the complete graph 
and this matches exactly the trivial upper bound
in the worst case.
However, Theorem~\ref{theo.lb2} shows that for this model the
for {\em almost all} labeled graphs asymptotically 50\% of 
this number of bits of total routing information 
is both necessary and sufficient.

Even if stretch factors between $1$ and $2$ are allowed, the next theorem
shows that
$\Omega(n^2 \log n)$ bits are necessary to represent the routing scheme
in the worst case.
\begin{theorem}\label{theo.aglb2}
For routing with stretch factor $< 2$ in labeled
graphs where relabeling is not
allowed 
($\alpha$),
there exist graphs on $n$ nodes (almost $(n/3)!$ such
graphs)
where the local routing function must be stored in at least
$(n/3) \log n - O(n)$ bits per node at $n/3$ nodes
(hence the complete routing scheme requires at least
$(n^2/9) \log n - O(n^2)$ bits to be stored).
\end{theorem}

	\begin{figure}
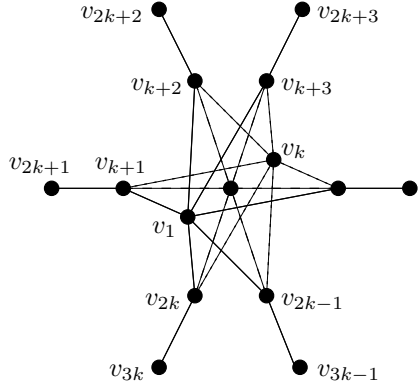

	\begin{center}
	\input stretch2.pstex_t
	\end{center}
	\caption{Graph $G_k$.}\label{fig-stretch2}
	\end{figure}

\begin{proof}
Consider the graph $G_k$ with $n=3k$ nodes depicted in
Figure~\ref{fig-stretch2}.
Each node $v_i$ in $v_{k+1},\ldots,v_{2k}$ is connected to
$v_{i+k}$ and to each of the nodes $v_1,\ldots,v_k$.
Fix a labeling of the nodes $v_1,\ldots,v_{2k}$ with labels from
$\{1,\ldots,2k\}$.
Then any labeling of the nodes $v_{2k+1},\ldots,v_{3k}$ with labels from
$\{2k+1,\ldots,3k\}$ corresponds to a permutation of $\{2k+1,\ldots,3k\}$
and vice versa.

Clearly, for any two nodes $v_i$ and $v_j$ with $1 \le i \le k$ and
$2k+1 \le j \le 3k$, the shortest path from $v_i$ to $v_j$ passes through
node $v_{j-k}$ and has length $2$, whereas any other path from $v_i$ to $v_j$
has length at least $4$. Hence any routing function on $G_k$
with stretch factor $< 2$ routes such $v_j$ from $v_i$ over
the edge $\edge{v_i}{v_{j-k}}$.
Then at each of the $k$ nodes $v_1,\ldots,v_k$ the local routing
functions
corresponding to any two labelings of the nodes $v_{2k+1},\ldots,v_{3k}$
are different.
Hence each representation of a local routing function at the $k$ nodes $v_i$,
$1 \le i \le k$, corresponds one-one to a permutation of $\{2k+1,\ldots,3k\}$.
So given such a local routing function we can reconstruct the permutation
(by collecting the response of the local routing function for
each of the nodes $k+1,\ldots,3k$ and grouping all pairs reached over the same
edge).
The number of such permutations is $k!$.
A fraction at least $1-1/2^k$ of such permutations $\pi$ has
Kolmogorov complexity $C(\pi) = k \log k - O(k)$~\cite{LiV93}.
Because $\pi$ can be reconstructed given any of the $k$ local routing
functions, these $k$ local routing functions each must have Kolmogorov
complexity $k \log k - O(k)$ too.
This proves the theorem for $n$
is a multiple of 3.
For $n = 3k-1$ or $n = 3k-2$ we can use $G_k$, dropping $v_k$ and $v_{k-1}$.
Note that the proof only requires that there be no relabeling;
apart from that the direct neighbors of a node may be known and 
ports may be reassigned.

By the above calculation there are at least $(1-1/2^{n/3})(n/3)!$
labeled graphs on $n$ nodes for which the theorem holds.
\qquad
\end{proof}

Our last theorem shows that for full information
shortest path routing schemes on Kolmogorov random graphs one cannot do better
than the trivial upper bound.
\begin{theorem}\label{th.fullinfo}
For full-information shortest path routing on
$o(n)$-random labeled gra\-phs on $n$ nodes 
where relabeling is not allowed ($\alpha$),
the local routing function
occupies at least $n^2/4 - o(n^2)$ bits for every node
(hence the complete routing scheme requires at least $n^3/4 - o(n^3)$
bits to be stored).
\end{theorem}
\begin{proof}
Let $G$ be a graph on nodes $\{1,2, \ldots, n\}$
satisfying Eq.~(\ref{eq.KG}) with $\delta (n) = o(n)$. Then we know that $G$
satisfies Lemmas~\ref{lem.degree}, \ref{lem.diam}. Let $F(u)$
be the local routing function of node $u$ of $G$, and let $|F(u)|$
be the number of bits used to encode $F(u)$. Let $E(G)$ be the standard
encoding of $G$ in $n(n-1)/2$ bits as in Def.~\ref{def.gc}.
We now give another way to describe $G$ using some
local routing function $F(u)$.
\begin{romannum}
\item
A description of this discussion in $O(1)$ bits.
\item
A description of $u$ in $\log n$ bits.
(If it is less pad the description with 0's.)
\item
A description of the presence or absence of edges between $u$ and the
other nodes in $V$ in $n-1$ bits.
\item
A description of $F(u)$ in $|F(u)| + O(\log |F(u)|)$ bits
(the logarithmic term to make the description self-delimiting).
\item
The code $E(G)$ with all bits deleted
corresponding to the presence or absence of edges between
each $w$ and $v$ such that $v$ is a neighbour of $u$ and $w$ is not a neighbour
of $u$.
Since there are at least $n/2-o(n)$ nodes $w$ such that
$\edge{u}{w} \notin E$
and at least $n/2-o(n)$ nodes $v$ such that
$\edge{u}{v} \in E$,
by Lemma~\ref{lem.degree}, this saves
at least $(n/2-o(n))^2$ bits.
\end{romannum}
From this description we can reconstruct $G$, given $n$,
by reconstructing the bits corresponding to the deleted edges
from $u$ and $F(u)$ and subsequently inserting them in the
appropriate positions to reconstruct $E(G)$. We can do so
because $F(u)$ represents a full information
routing scheme implying that
$\edge{v}{w} \in E$ iff
$\edge{u}{v}$ is among the edges used to route from
$u$ to $w$.
In total this new description has
\[ n(n-1)/2 + O(\log n) + |F(u)|
- n^2/4 + o(n^2) \]
which must be at least $n(n-1)/2 -o(n)$ by Eq.~(\ref{eq.KG}).
We conclude that $|F(u)| = n^2/4 - o(n^2)$, which proves the theorem.

Note that the proof only requires that there be no relabeling;
apart from that the direct neighbors of a node may be known and
ports may be reassigned.
\qquad
\end{proof}

\section{Average Case}
What about the average cost, taken over all labeled graphs
of $n$ nodes, of representing a routing scheme for graphs over $n$ nodes?
The results above concerned precise overwhelmingly large
fractions of the set of all labeled graphs.
The numerical values of randomness deficiencies and
bit costs involved show that these results are actually
considerably stronger than the corresponding average case
results which are straightforward.
\begin{definition}
\rm
For each labeled graph $G$, 
let $T_S (G)$ be the minimal total number of bits used 
to store a routing
scheme of type $S$ (where $S$ indicates shortest path routing,
full-information routing, and the like.).
The {\em average} minimal total number of bits to store a routing scheme
for $S$-routing over labeled graphs on $n$ nodes is
$\sum T_S(G)/2^{n(n-1)/2}$ with the sum taken over all graphs
$G$ on nodes $\{1, 2, \ldots, n\}$. (That is, the uniform average
over all the labeled graphs on $n$ nodes.)
\end{definition}

The results on Kolmogorov random graphs above have the following
corollaries. The set of $(3 \log n)$-random graphs
constitutes
a fraction of at least $(1-1/n^3)$ of the set of all graphs
on $n$ nodes. The trivial upper bound on the minimal total number
of bits for all routing functions together is $O(n^2 \log n)$ for shortest
path routing on all graphs
on $n$ nodes (or $O(n^3)$ for full-information shortest path
routing). Simple computation 
shows that the average total number
of bits to store the routing scheme for graphs of $n$ nodes is
(asymptotically and ignoring lower order of magnitude terms as in
Table~\ref{tab.results}):

\begin{remunerate}
\item $ \leq 3n^2$ for shortest path routing in model IB \mor II
	(Theorem~\ref{theo.ub1});
\item $ \leq 6 n \log^2 n$ for shortest path routing in model II \mand $\gamma$
	where the average is taken over the initially
	labeled graphs on $n$ nodes with labels in $\{1,2, \ldots , n\}$
	{\em before} they were relabeled with new 
	and longer labels giving routing information
	(Theorem~\ref{theo.ub5});
\item $ \leq 38 n \log n$ for routing with any stretch factor $s$  for $1 < s < 2$
	in model II (Theorem~\ref{theo.ub4});
\item $ \leq n \log \log n$ for routing with stretch factor 2 in model II
	(Theorem~\ref{theo.ub2});
\item $O(n)$ for routing with stretch factor $6 \log n$ in model II
	(Theorem~\ref{theo.ub3} with $c=3$);
\item $ \geq n^2/2$ for shortest path routing in model $\alpha$
	(Theorem~\ref{theo.lb1});
\item $ \geq n^2 /32$ for shortest path routing in model IA and IB
	(under all relabeling conventions,
	Theorem~\ref{theo.lb1-a});
\item $ = (n^2/2) \log n$ for shortest path routing in model IA \mand $\alpha$
	(Theorem~\ref{theo.ub6} and Theorem~\ref{theo.lb2});
\item $= n^3 /4$ 
 for full information shortest path routing in mo\-del
	$\alpha$ (Theorem~\ref{th.fullinfoub} and Theorem~\ref{th.fullinfo}). 
\end{remunerate}

\section{Conclusion}
The space requirements for compact routing for almost all labeled graphs
on $n$ nodes, and hence for the average case of all graphs on $n$ nodes,
is conclusively determined in this paper. We introduce a novel
application of the incompressibility method.
The next question arising in compact routing is the following.
For practical purposes
the class of all graphs is too broad in that most graphs have
high node degree (around $n/2$). Such high node degrees are unrealistic
in real communication networks for large $n$. The question arises
to extend the current treatment to almost all graphs on $n$ nodes of maximal
node degree $d$ where $d$ ranges from $O(1)$ to $n$. Clearly,
for shortest path routing $O(n^2 \log d)$ bits suffice,
and \cite{GavP96} showed that
for each $d < n$
there are shortest path routing schemes that require
a total of $\Omega (n^2 \log d)$ bits to be stored
in the worst case for some graphs with maximal degree $d$,
where we allow that nodes are relabeled by permutation and
the port-assignment may be changed (IB \mand $\beta$).
This does not hold for average routing since by our Theorem~\ref{theo.ub1}
$O(n^2)$ bits suffice for $d= \Theta(n)$. (Trivially, $O(n^2)$ bits 
suffice for routing in every graph with $d=O(1)$.)
We believe it may be possible to show by an extension of our method that
$\Theta (n^2)$ bits (independent of $d$) are necessary and sufficient
for shortest path routing in almost all graphs of maximum node degree $d$,
provided $d$ grows unboundedly with $n$.

Another research direction is to resolve the questions
addressed in this paper for Kolmogorov random unlabeled graphs
in particular with respect to the free relabeling model
(insofar as they do not follow a fortiori from the results presented here).

\section*{Acknowledgements}

We thank Jan van Leeuwen, Evangelos
Kranakis and Danny Krizanc for helpful discussions, and
the anonymous referees for comments and corrections.

\end{document}